# ELECTRON COOLING EXPERIMENTS IN CSR*

Xiaodong Yang[#], Guohong Li, Jie Li, Xiaoming Ma, Lijun Mao, Ruishi Mao,

Tailai Yan, Jiancheng Yang, Youjin Yuan , IMP, Lanzhou, 730000, China

Vasily V. Parkhomchuk, Vladimir B. Reva , BINP SB RAS, Novosibirsk, 630090, Russia

*Abstract*

The six species heavy ion beam was accumulated with the help of electron cooling in the main ring of Cooler Storage Ring of Heavy Ion Research Facility in Lanzhou(HIRFL-CSR), the ion beam accumulation dependence on the parameters of cooler was investigated experimentally. The 400MeV/u $^{12}C^{6+}$ and 200MeV/u $^{129}Xe^{54+}$ was stored and cooled in the experimental ring CSRe, the cooling force was measured in different condition.

## INTRODUCTION

Heavy Ion Research Facility of Lanzhou(HIRFL)[1] is an accelerators complex with multi-purpose, its research field includes radioactive ion beam physics, heavy ion physics, high energy density physics, super-heavy elements synthesis, atomic physics, and cancer therapy. It consists of two cyclotrons, SFC and SSC, two synchrotrons, CSRm and CSRe. It can provide the ion beam with energy range from 10 MeV/u to 1GeV/u. The ion beam delivered from SFC or SSC was injected into CSRm, after accumulation with the help of electron cooling, and acceleration, and then delivered to cancer therapy terminal and other experimental terminals, or injected into CSRe. In CSRe, ion beam was cooled by electron cooling device, and various physics experiments were completed in this ring. The ion beam with higher energy in CSRe was stripped, and higher charge state ion beam will be decelerated to lower energy, in the case of low energy of higher charge state ion beam, atomic physics experiments will be performed in CSRe.

## ELECTRON COOLING DEVICES

The electron cooling devices was equipped in each ring of CSR, the purpose of electron cooling in CSRm is ion beam accumulation, the cooler was adapted as the way to increase the stored particle number in CSRm, continuous electron cooling is applied to the stored ion beam for compensation of the heating by an internal gas jet target in CSRe, the most important feature is the ability to cool ion beam to highest quality for experiments with stored highly charged ions.

In CSRm, the electron cooling device plays an important role in the heavy ion beam accumulation at injection energy. The new state-of-the-art electron cooling device was designed and manufactured in the collaboration between BINP and IMP, it has three distinctive characteristics, namely high magnetic field parallelism in cooling section, variable electron beam profile and electrostatic bending in toroids.

Continuous electron cooling is applied to the stored ion beam for the compensation of the heating by various scattering in CSRe. The most important thing is the ability to cool the ion beams to the highest quality for physics experiments with stored highly charged ions. The electron cooling devices of HIRFL-CSR were reported in many conferences[2],[3],[4],[5],[6],[7],The previous results have been given in the COOL05-P02[8], COOL07-TUM1I02[9] and COOL09-FRM1MCIO02[10].

## ION BEAM ACCUMULATION IN CSRM

In order to demonstrate the performance of HIRFL accelerators complex, and satisfy the requirements of different physics experiments, ion beam with different energy, different charge state were accumulated with the help of electron cooling in CSRm. During accumulation, two injection modes were applied, in the case of lighter ion beam, stripping injection was adapted, for heavier ion beam, repeated multi-turn injection was performed. Due to the injection beam intensity, ion beam was delivered by different injector, SFC or SSC. In the case of fixed energy, choose proper injection interval, partially hollow electron beam, the direction and position of electron beam and ion beam matched well, the maximum accumulation results can be achieved.

*Commissioning procedure*

The CSR commissioning procedure was described as following steps:
- Obtaining high transportation efficiency in beam line and maximal beam intensity at injection point.
- Correcting the position and angle of ion beam at the injection point, Obtaining the maximal injection intensity at first Faraday cup in the ring.
- Correcting the closed-orbit globally and locally, specially in the region of electron cooler, correcting work-point (without electron beam and with electron beam), improving ion beam lifetime in the ring.
- After turned on the cooler magnetic field, compensating the influence of cooler magnetic field, correcting the position and angle of ion beam entering cooler.
- Fine tuning the energy of electron, after observe accumulation, optimizing the electron beam current and the profile, improve the lifetime of ion beam in the present of electron beam.
- Optimizing injection interval, bump amplitude and time constant.
- Optimizing the ramping data, proper time setting of trigger for RF and kicker, acceleration to high energy

___________________________________________
*Work supported by The National Natural Science Foundation of China, NSFC(Grant No. 10975166, 10905083, 10921504)
[#]yangxd@impcas.ac.cn

and extracted from CSRm.

Table 1: Accumulation results in CSRm

| Ion | $E_{inj}$ MeV/u | Injector | M | $I_{inj}$ μA | $\Delta T_{inj}$ s | $I_{single}$ μA | $I_{10\,sec}$ μA |
|---|---|---|---|---|---|---|---|
| $^{12}C^{6+}$ | 7.09 | SFC | ST | 12 | 1.0 | 167 | 700 |
| $^{12}C^{4+}$ | 7.1 | SFC | MI | 6 | 1.0 | 20 | 105 |
| $^{36}Ar^{18+}$ | 21.7 | SFC+SSC | MI | 4 | 0.35 | 6 | 250 |
| $^{129}Xe^{27+}$ | 2.9 | SFC | MI | 3 | 0.35 | 6.5 | 70 |
| $^{12}C^{5+}$ | 8.26 | SFC | MI | 3 | 0.9 | 11 | 70 |
| $^{78}Kr^{28+}$ | 4.04 | SFC | MI | 2.4 | 0.2 | 5 | 80 |

Some accumulation results were summarized in Table. 1. The $I_{inj}$ indicated the beam intensity in the beam line, it was limited by the capability of injector. The $\Delta T_{inj}$ is the injection interval, it depends on the transverse cooling time, the electron cooling parameters for different ion beam with different energy. $I_{single}$ is the average ion intensity in one standard multi-turn injection, it depends on the ion beam pulse length of injector. $I_{10sec}$ is the final ion beam intensity accumulated in 10 sec, the cycle for ion beam accumulation. A typical DCCT signal of injection results of $^{12}C^{6+}$ was displayed in Fig. 1.

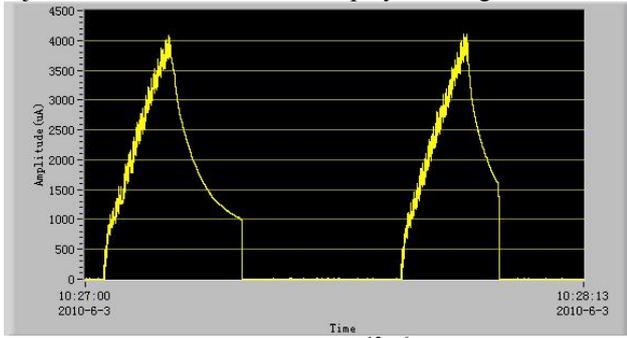

Figure 1: DCCT signal during $^{12}C^{6+}$ accumulation in CSRm.

*Stripping injection*

Firstly, the 7 MeV/u $^{12}C^{4+}$ was injected into CSRm from the small cyclotron SFC through a stripping foil with thickness of 15 μg/cm² placed in the first dipole of the ring, the average pulse intensity was about 8.4 μA in the injection line. In the absence of magnetic field of the electron cooler, the single-turn stripping injection beam was tested in CSRm with bumping orbit, the stored beam signal was observed from BPM signal, the closed orbit correction was done roughly, the machine parameter such as work-point was measured and tuned, and acceleration attempted. The average particle number of stored $^{12}C^{6+}$ was about $4.7\times10^8$ in one standard multi-turn injection. With the help of electron cooling of partially hollow electron beam, $2.5\times10^9$ particles were accumulated in the ring after 10 times injection in 10 seconds.

Due to intensive commissioning of Carbon ion beam in CSRm, some accumulation results were listed in Table. 2. The repeated multi-turn injection of $^{12}C^{6+}$ was attempted. $^{12}C^{4+}$ directly extracted from ECS ion source, and stripped into $^{12}C^{6+}$ in the beam line. The accumulation rate was about 3.5. In the later commissioning, the stripping injection was applied. The accumulation rate increased to 5.8 in the second row in Table 2. After fine tune the electron cooling parameters, especially the position and angle between the ion and electron beams, shortened the injection interval, 2100μA $^{12}C^{6+}$ was obtained in 10 seconds. According to the gain formula, gain=repetition frequency×lifetime. It is helpful to increase the injection number, and to shorten the injection interval. In the latest commissioning, near $1.8\times10^{10}$ particles were accumulated in 10 seconds.

Table 2: Accumulation parameters of ion beam

| Ion | $E_{inj}$ MeV/u | Injector | M | $I_{inj}$ μA | $\Delta T_{inj}$ s | $I_{single}$ μA | $I_{10\,sec}$ μA |
|---|---|---|---|---|---|---|---|
| $^{12}C^{6+}$ | 7.09 | SFC | MI | 4.3 | 0.5 | 12.5 | 150 |
| $^{12}C^{6+}$ | 7.09 | SFC | ST | 12 | 1.0 | 104 | 700 |
| $^{12}C^{6+}$ | 7.09 | SFC | ST | 11 | 0.5 | 260 | 2100 |
| $^{12}C^{6+}$ | 7.04 | SFC | ST | 8.4 | 0.25 | 400 | 4000 |

The two components accumulation fitting comparing with the experimental data was demonstrated in Fig. 2. From the results, the average intensity in one standard multi-turn injection was about 350μA, one part of ion beam decayed with the lifetime of 6 seconds, the other part of ion beam decayed with 0.35 seconds. In this condition, the measured work-point was 3.612/2.657, the improper work-point was the reason of short lifetime and fast decay of ion beam.

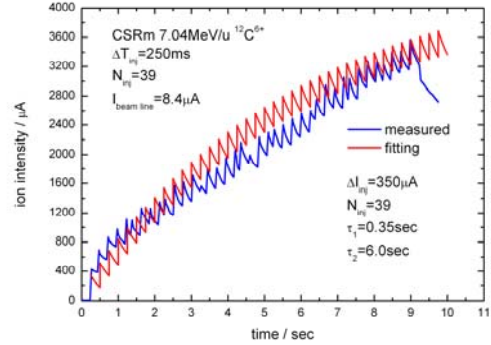

Figure 2: Two components decay fitting comparing with experimental data.

*Repeated multi-turn injection*

At the end of the transfer line, a magnetic septum and an electrostatic septum inflector guide the beam parallel to the ring orbit; four in-dipole coils create a DC bump of 50 mm amplitude at the electrostatic septum. For multi-turn injection four fast bump magnets produce a time dependent bump orbit to fill the horizontal acceptance of the ring.

After repeated multi-turn injection, the emittance of ion beam will be close to the transverse acceptance of the ring. And the radius of ion beam will be 3.8 cm in the cooling section. The ion beam is completely surrounded by the electron. The accumulation was improved in the case of bigger expansion factor.

*Accumulation optimization experiments*

The accumulation rate subjects to the cooling time and injection repetition rate. It is determined by the electron beam parameters and injected ion beam stability. The optimum time interval between the two adjacent multi-turn injections corresponds to the transverse cooling time of ion. After observed the accumulation, the parameters related to accumulation were optimized experimentally. The dependence of accumulated ion intensity in 10 seconds on the injection interval was shown in Fig. 3. From this result, in the case of the injection interval with 0.25 seconds, the maximal ion intensity was obtained.

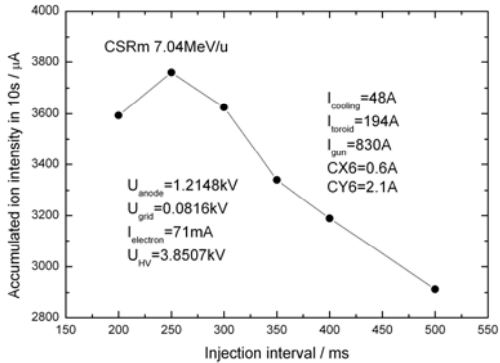

Figure 3: Accumulated ion intensity in 10s as a function of the injection interval.

The position and angle between ion and electron beams in the cooling section determined the cooling force and time. The accumulated ion intensity in 10s as a function of the current in cooling section correctors CX6 and CY6 were illustrated in Fig.4 and Fig. 5, the electron beam angles in horizontal and vertical direction with respect of ion beam were changed by these correctors. It's obvious that perfect alignment is helpful for obtaining maximum ion intensity. Due to the improper orbit of ion beam in the cooling section, the results were not as usual parabolic curve. The current regulation range of correctors was limited by the aperture of electrostatic bending plates in the toroids of cooler because of the condition of bigger expansion factor. Excessive regulation caused the fast increasing of the load current of high voltage system.

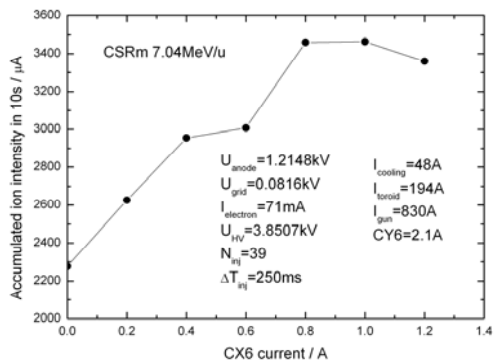

Figure 4: Accumulated ion intensity in 10s as a function of the current in cooling section corrector CX6.

The accumulated ion intensity in 10 s was measured as a function of the ratio $U_{grid}/U_{anode}$ of electron gun at different electron current as presented in Fig. 6. It is clear that optimum accumulation happens in the partially hollow electron beam, the ratio $U_{grid}/U_{anode}$ is close to 0.1. In this case, the central density is less than the edge one in the electron beam. The energy of electron beam was fixed, the potential drop caused by the space charge of electron beam with a different profile was not taken into account in the experiments.

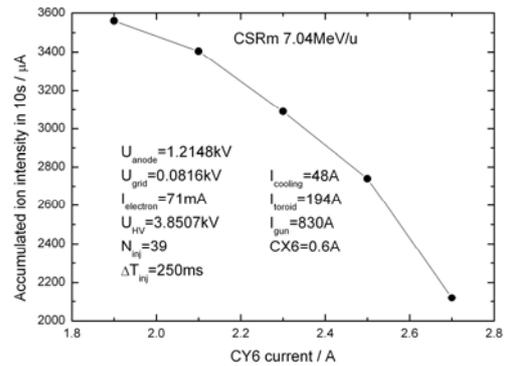

Figure 5: Accumulated ion intensity in 10s as a function of the current in cooling section corrector CY6.

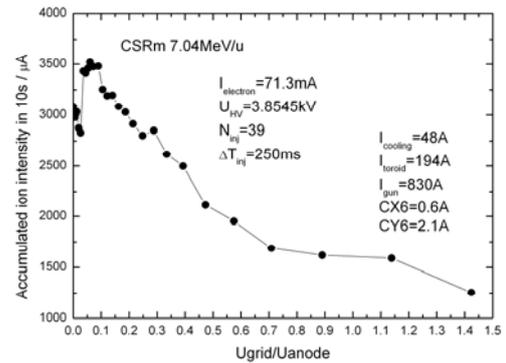

Figure 6: Accumulated ion intensity in 10s as a function of the ratio $U_{grid}/U_{anode}$ of electron gun.

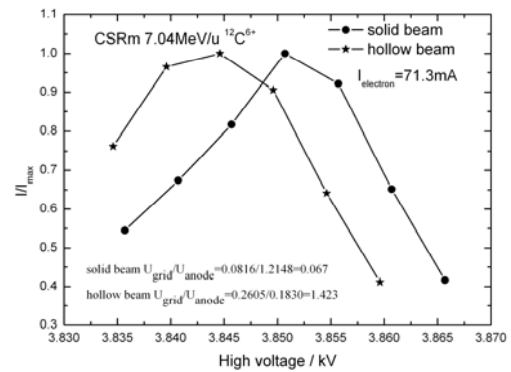

Figure 7: Accumulated ion intensity in 10s as a function of the energy for solid and hollow electron beam.

The optimal electron beam energy in the case of different electron beam profile was shown in Fig. 7. The intensity of ion beam was normalized by the maximum value. The electron beam current was kept fixed in the experiments. It is clearly shown the effects of space charge for the solid and hollow electron beam.

## ION BEAM COOLING IN CSRE

The ion beam of $^{12}C^{6+}$ was injected into CSRe from CSRm after acceleration from 7MeV/u to 200MeV/u. The magnetic field of electron cooler was set as a quarter of the maximum value. In this case, the magnetic field in the cooling section was 0.0385 T, and the electron beam current was set as 300 mA. After the global and local orbit correction and regulation of the electron beam angle and position in cooling section, the cooling was observed, but the cooling process was not fast enough. It could be caused by the poor quality of the high energy electron beam confined by a weak magnetic field in toroid, where additional transverse temperature was introduced. After increasing the magnetic field in toroid, increasing the electron beam current, improving the angle between ion beam and electron beam, about 14 seconds longitudinal cooling time was obtained. Then the ion beam was accelerated to 400MeV/u, the corresponding high voltage of electron cooler was 220kV, and the electron beam current achieved to 1A, a reasonable momentum spread of $3.2\times10^{-5}$ was measured in the case of 600μA ion beam in CSRe. Some results have been reported in COOL'09[10].

In order to explore the minimum momentum spread, the ion beam was cut in the beam line before CSRm by means of changing the current of last quadrupole, it resulted in the DCCT in CSRm had no signal obviously. In this case, the particle number was less than $1.5\times10^{6}$ in CSRm, after acceleration, the particle number was less than $9\times10^{4}$ in CSRe, but the Sckottky monitor had the clear signal, the minimum momentum spread was measured as $1.35\times10^{-5}$ demonstrated in Fig. 8. Generally, the measured minimum momentum spread was limited by the detection technique, stability of dipole power supply, the stability of high voltage power supply of cooler, and also the particle number stored in the ring.

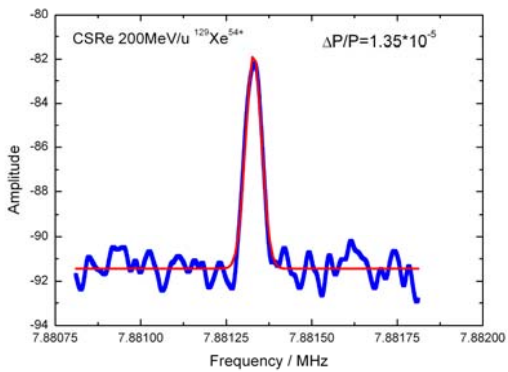

Figure 8: Measured Schottky signal of $^{129}Xe^{54+}$ in CSRe in the case of low ion beam intensity.

From the experience of CSRe cooler, the stability of dipole power supply and high voltage power supply of cooler should be improved in future. The detection technique should be upgraded to measure the momentum spread in the case of low ion intensity, and precision calibration should be done for determine the stored particle number.

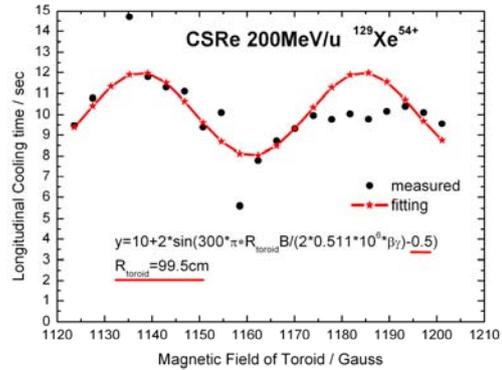

Figure 9: Longitudinal cooling time measured as a function of the magnetic field in toroid.

In the internal gas target experiments, the 200MeV/u $^{129}Xe^{27+}$ was stripped into $^{129}Xe^{54+}$ before entry of CSRe, and stored in CSRe. The longitudinal cooling time was measured as a function of magnetic field of toroid, the results was presented in Fig.9, the angle and position of the electron beam change during only regulating the current of toroid was not taken into account. From this result, one can see the dependence of cooling time on the magnetic field of toroid, it is good agreement with the results of GSI[12].

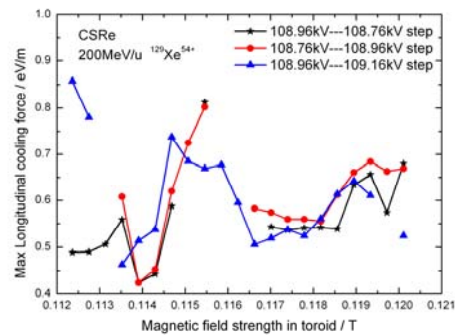

Figure 10: Maximum longitudinal cooling force as a function of the magnetic field in toroid in the case of different electron beam energy steps.

The influence of the magnetic field in toroid on the longitudinal cooling force for 200 MeV/u $^{129}Xe^{54+}$ beam was studied experimentally, longitudinal cooling forces were obtained by means of the electron energy-step method[11]. The cooling force varies with magnetic field in toroid was shown in Fig.10. The different line represents the different electron energy step. In the case of high energy, the influence of magnetic field on the electron transverse energy should be paid sufficient attention. The experimental results were in partially agreement with the experiment results in GSI[12].

From those results after the electron beam was switched off, the stability of high voltage system of CSRe cooler was derived. The data of high voltage (HV) was recorded as one file. Fig. 11 reveals the results of stability of HV of

CSRe cooler, the black line indicated the temperature from the monitor in the collector, the red line shown the signal from divider resistor 2 of HV system, one can find that as the change of temperature during 10 days, the output of high voltage changed with the temperature. From this point of view, the stability of HV system of CSRe cooler should be improved in the future.

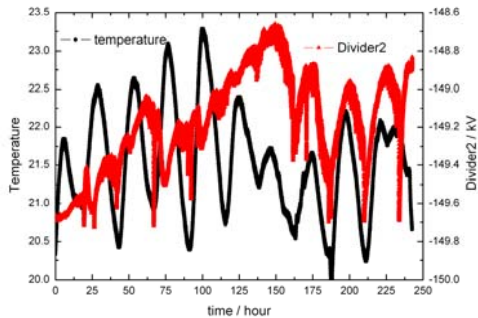

Figure 11: The temperature and divider2 voltage of CSRe cooler as a function of the time.

## UPGRADE AND IMPROVEMENT

Due to the ion orbit was not in the centre of the compensation solenoid at the ends of electron cooler, when the current of compensation solenoid increased, the ion beam was not pass through the solenoid properly. It limited the increasing the working magnetic field in the cooler.

Various lifetimes were observed in different condition, it was caused by the improper setting of work-point of ring, in additional, the correction of closed-orbit was not carefully done during the operation, in this case the ion beam lost fast.

The work-point of ring varying with electron beam current of cooler was found in operation. This should be investigated experimentally in the future.

A special system of electron energy modulation was installed and tested in CSRm cooler, the energy of electron can be modulated in negative or positive deviation to the optimal value, the amplitude, pulse length and repetition frequency can be change according to the experiment requirements.

## SUMMARY

- A few species heavy ion beam with different injection energy was cooled, accumulated and accelerated in CSRm.
- 400MeV/u $^{12}C^{6+}$ and 200MeV/u $^{129}Xe^{54+}$ was cooled with internal target in CSRe.
- Two cooling devices come into routine operation.
- Electron beam energy modulation system was installed and tested in CSRm cooler.
- Coolers were ready for physics experiments.